# A new approach (extra vertex) and generalization of Shoelace Algorithm usage in convex polygon (Point-in-Polygon)

Rakhmanov Ochilbek
*Department of Computer Science*
*Nile University of Nigeria*
Abuja, Nigeria
ochilbek.rakhmanov@nileuniversity.edu.ng

*Abstract-* **In this paper we aim to bring new approach into usage of Shoelace Algorithm for area calculation in convex polygons on Cartesian coordinate system, with concentration on point in polygon concept. Generalization of usage of the concept will be proposed for line segment and polygons. Testing of new method will be done using Python language. Results of tests show that the new approach is more effective than the current one.**

## I. INTRODUCTION

Point-in-polygon (PiP) is one of the fundamental operations of Geographic Information Systems. Yet, nowadays this concept is also getting big attention in graphical programming, mobile game programming, and other fields. The "point-in-polygon" problem is defined as: "With a given polygon P (polygon ABCDE in this case) and an arbitrary point F (Fig.1), determine whether point F is enclosed by the edges of the polygon".

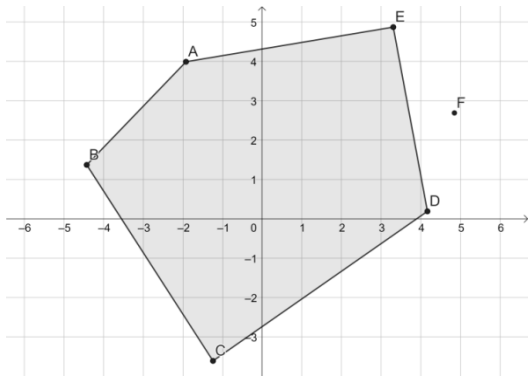

Fig. 1. A polygon ABCDE with point F outside of its premises.

This example doesn't appear to be difficult to answer, but in real life application it may get complex when polygon has high number of edges and it may need to do this calculation for many points. Yet, it is also gaining high importance in modern game programming. For instance, during programming a game where opponents attack each other's premises with missiles, it is necessary to calculate whether the coordinate point where the missile landed is inside or outside of opponent's polygonic premises.

A number of algorithms can be applied to do this calculation. Some known methods are; sum of angles, ray shooting, signed angle method, sum of area and some others [3]. Different algorithms lead to different running efficiencies, but all lead to correct answer.

Of course, the shape of polygon is also important in application of those methods. Some will show high efficiency only for convex polygons, but may not do same for concave polygons. While others can deal with concave polygons, but time complexity may increase.

This paper aims to bring a new approach to "Sum of area" method in a way of calculation and coding. Generalization of usage of this method will be proposed as well. Python will be used as a medium for tests.

The well-known shoelace algorithm (shoelace formula) is used to calculate the area related problems in polygons. The algorithm is called so, since it looks like a shoelace during cross product calculation of matrix. It was proposed by Gauss, in 1795 [2]. The basic idea be-hind the algorithm is to divide the polygon into triangles and calculate sum of area of all formed triangles. Fig.2 shows similarity between matrix multiplication of vertices and real shoelace tying.

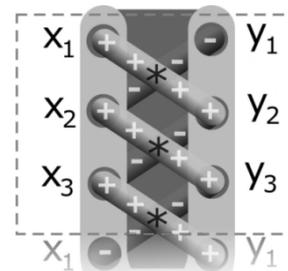

Fig. 2. Representation of shoelace algorithm

Actually, sum of area method is also using the same idea used in shoelace algorithm. It divides the polygon into many triangles, with one fixed vertex and calculates the area of all formed triangles. The approach this paper proposing is still using triangle method, but not with one fixed point vertex.

We should underline this paper's methodology will be more concentrated on convex polygons. But for the case of concave polygons, and possible errors or clashes, some observations will be done. In other words, if it is case for concave polygon, the generalization for line segment or polygon may not suit completely, but one can overcome with some additions.

## II. REVIEW OF RELATED LITERATURE

As the necessity for Point-in-Polygon concept was growing in mid 1980s, several studies have been conducted. Of course all studies were based on coding as well. Just proposing method without assuming the time complexity during programming was not a point. Recent studies are more concentrated on edition of existing methods or edition of coding of proposed methods.

As it stands, the most basic method is Sum of Area method, as we mentioned above, this method divides the polygon into triangles and calculates the total area of all triangles using shoelace algorithm. Dave Rusin proposed the method in 1995[3].

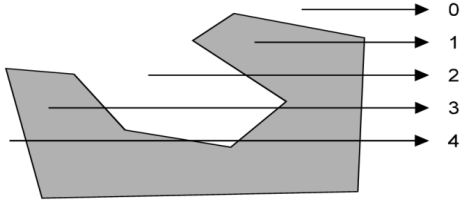

Fig. 3. Ray shooting concept represantation.

One of most popular algorithms is Ray Shooting method. Randolph Franklin converted Ray Shooting method into 7 line C code, after several trials with his doctoral students. The algorithm is simple; from any chosen point, you go through a ray to any side of a polygon. If the ray intersects the polygon in odd number times, than it is inside of it, for even number it is oppose, it is outside of it. Fig.3 is showing the Ray Shooting method application.

One of other remarkable methods is Sum of Angles method. The algorithm is to compute the sum of the angles made between the test point and each pair of points making up the polygon. If this sum is $-2\Pi$ then the point is an interior point, if 0 then the point is an exterior point. The method and coding was proposed by Philippe Reverdy [5].

There are other methods as well. We mentioned above that recent studies are also conducted to compare them and their complexity. One of such studies is done by C.W. Huang et al, where some methods are compared in details [1]. In conclusion of this paper, as it was expected, they didn't propose any algorithm as front runner, instead they concluded that algorithms behave relatively to the shape of polygon or complexity of calculation.

### III. METHODOLOGY

#### A. Shoelace Algorithm

Suppose the polygon P has vertices $(x_1, y_1), (x_2, y_2), ..., (x_n, y_n)$, listed in clockwise order. Then the area of polygon P is:

$$A = \frac{1}{2}\left|(x_1 y_2 + x_2 y_3 + ... + x_n y_1) - (y_1 x_2 + y_2 x_3 + ... + y_n x_1)\right|$$

Alternatively, it can be shown as in 3.1:

$$A = \frac{1}{2}\left|\sum_{i=1}^{n}(x_i y_{i+1} - x_{i+1} y_i)\right| \quad (3.1)$$

Same formula can be presented in matrix form, clearly shows how the multiplication is formed.

$$A = \frac{1}{2}\left|\sum_{i=1}^{n}\det\begin{pmatrix} x_i & x_{i+1} \\ y_i & y_{i+1} \end{pmatrix}\right| = \frac{1}{2}\begin{vmatrix} x_1 & y_1 \\ x_2 & y_2 \\ \vdots \\ x_n & y_n \end{vmatrix} \quad (3.2)$$

Fig.4 clearly describes how the triangulation is applied into polygon to calculate the area. So area of polygon P in Fig.4 can be calculated as:

$$A(ABCDEFG) = A(ABC) + A(ACD) + A(ADE) + \\ + A(AEF) + A(AEG)$$

#### B. Point in polygon

The PiP concept can be applied to any size of polygon. Yet, to keep it simple for understanding and visible, polygon P in Fig.4 will be used.

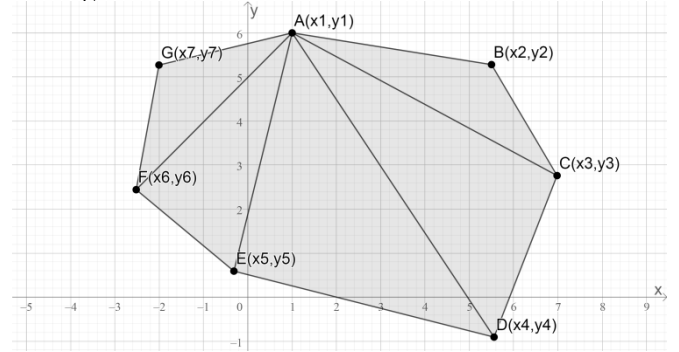

Fig. 4. Triangulation with fixed vertex

Assume that point H is inside of enclosed polygon and point K is outside. Fixing these points as one of triangle vertices, divide polygon into triangles with respect to each side of polygon. Fig.5 demonstrates the triangulation for point H, and Fig.6 for point K.

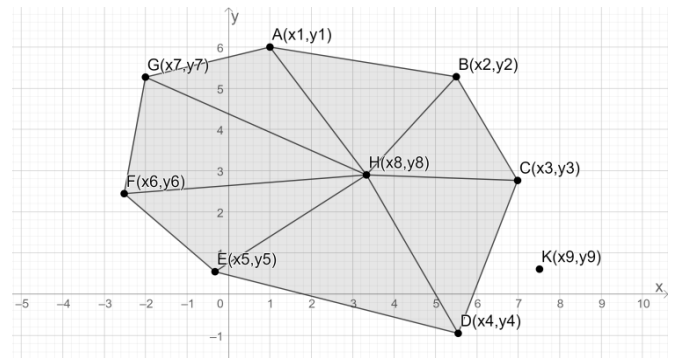

Fig. 5. Triangulation with point H

If we calculate total area of newly formed 7 triangles with fixed vertex H, it is clear that the area will be same with our original polygon P, with area of $A(ABCDEFG)$. No matter where the point H will be, inside the polygon, result will be same, even if the point H is on the line segment of one of the sides of the polygon.

In comparison to point H, point K will lead to completely different result. Fig.6 demonstrates that sum of all the newly formed 7 triangles will be bigger than the actual size of the polygon P. No matter how the K is selected, there will be always an excessive area formed outside of polygon.

This concept of methodology was presented and used till today [3].

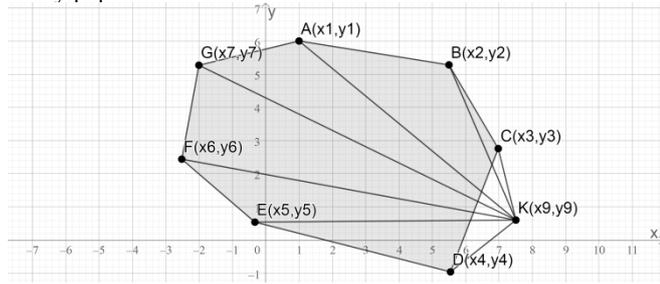

Fig. 6. Triangulation with point K.

*C. A new approach on point in polygon*

The new approach this paper is proposing is to evaluate point H, or K, as one of the points of the polygon listed in clockwise direction and forms a new polygon P1 with vertices ABCDHEFG, or P2 with vertices ABCKDEFG, respectively. This will definitely lead to less complex structure. We can call it *extra vertex* concept since we add one extra vertex to existing polygon.

Fig.7 and Fig.8 are representations on how to evaluate points as just next one in clockwise direction.

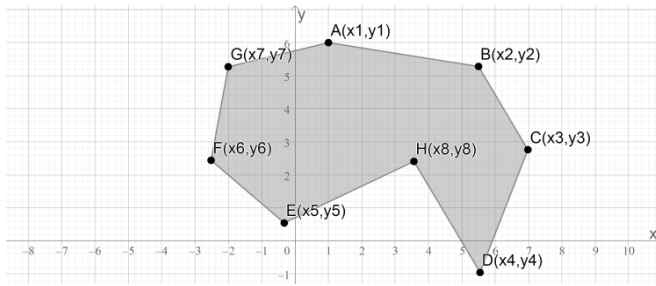

Fig. 7. Point H as a extra vertex.

It is clear to observe, from Fig.6, that when the point H is inside the polygon, the new formed polygon P1 (ABCDHEFG) will definitely have smaller area size than original polygon P (ABCDEFG).

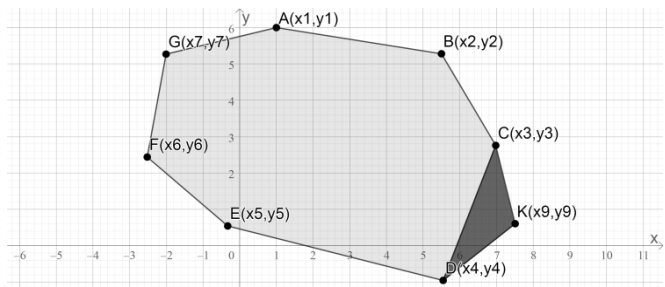

Fig. 8. Point K as a extra vertex.

Point H will push one of the sides of the polygon to inner side of the polygon; consequently, will decrease the area of the polygon, while point K will form extra triangle with one of the side, hence will increase the area of the polygon. As a result of comparison between areas of polygons P, P1 and P2 will give us results about whether the new point is inside of polygon region or outside of it. Table 1 shows results of comparison between the areas and conclusion with assumption a new point R is tested. P is area of original polygon, while PR is area of new formed polygon.

Table I

| Comparison | Conclusion |
|---|---|
| $A(P) > A(P_R)$ | R is inside of polygon P |
| $A(P_R) > A(P)$ | R is outside of polygon P |
| $A(P) = A(P_R)$ | R is on line of one of the edges of P |

*D. Generalization of the concept*

*1. Line Segment*

If it is needed to check whether a line segment lies in-side the polygon region, then it is sufficient to check if the beginning and ending point are both inside the polygon. This generalization can only contradict if the polygon is concave. In this case, it is necessary to check whether the line segment is intersecting with any sides of the polygon. Fig.9 demonstrates the test for line segment TS.

*2. Polygon*

To check whether a new given polygon is completely inside of the polygon P, just like in line segment condition, all the vertices of the polygon should be checked whether they are inside of P. Fig.9 demonstrates how all vertices of New Polygon (LJONM) are placed inside of P. If any of those vertices will drop out than it is clear that polygons are overlapping.

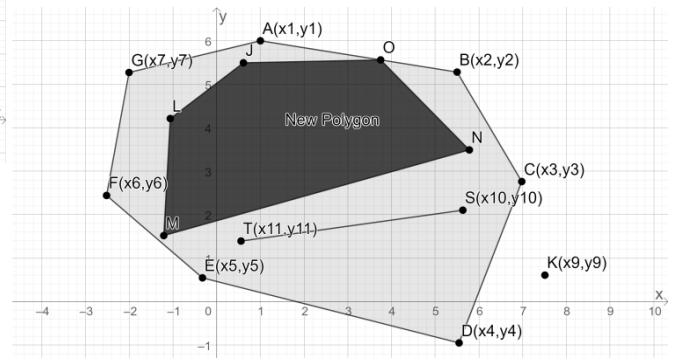

Fig. 9. Visualization of line segment and polygon in side of P.

*E. Comparison in programming*

Language to test the code is selected as Python and environment is chosen as Jupyter Notebook. Following steps are done:

*1. Assigning variables*

$x=[x_1,x_2,…,x_n]$ , $y=[y_1,y_2,…,y_n]$    *(x and y coordinates)*

k=number of vertices

Ps = area calculated with SLA

Pt= area calculated with triangulation

$x_f$, $y_f$ = fixed point for triangle calculation

$x_{out}$, $y_{out}$ = new points chosen outside of polygon

$x_{in}$, $y_{in}$ = new points chosen inside of polygon

*2. Shoelace algorithm and result of area calculation:*

```
def AreaSL(x,y,k):
    for i in range (0,k):
        Calculate area using (3.2)
    return float area (or double)
P=AreaSL(x,y,k)
```

*3. Triangulation of polygon and area calculation result*

```
def AreaT(x_f, y_f , x,y,k):
    for i in range (0,k):
        calculate area using (3.2), for
            3 vertices with one fixed (x_f, y_f)
    return float area (or double)
Pt=AreaT(x[i],y[i],x,y,k)          (#i is any edge on P)
```

It is necessary to underline that choosing the type of the result will affect this result. Round, float or double can be used. It depends how precise should be the area calculation.

*4. Assigning new points for test and appending them to existing list:*

```
x=x.append(x_new)
y=y.append(y_new)
```

*5. Testing new points with Shoelace algorithm, using extra vertex concept.*

```
x=x.append(x_in)
y=y.append(y_in)
k=k+1
if(AreaSL(x,y,k)>P): outside point
else: inside point
```

```
x=x.append(x_out)
y=y.append(y_out)
k=k+1
if(AreaSL(x,y,k)<P): inside point
else: outside point
```

*6. Testing new points with triangulation:*

```
if(AreaT(x_in, y_in ,x,y,k)>P): outside p
else: inside point
```

```
if(AreaT(x_out, y_out ,x,y,k)>P): outside p
else: inside point
```

*F. Analysis of algorithms( existing and extra vertex)*

Both of the algorithms (triangulation method and extra vertex method) loops for n times, hence both can be accepted as linear function. So time complexity for both is *O(n)*. But, constant *c* will vary for them. To test this we produced random polygon with 20 vertices and 180 vertices. Using *timeit* library in Python, we recorded run length of the algorithms. For given vertices we run algorithm for 100 times and recorded mean of that of all 100 outcomes. We did this test 20 times, so totally 20 different test means results were recorded. The following formulation was used to test time usage:

```
for i in range (0,20)
    for i in range (0,100)
        start time
        test with Extra V
        stop time
        value=start-stop
        record value in list
    average of values in list
    record avr value in new list
results of test
```

Same formulation for triangulation method:

```
for i in range (0,20)
    for i in range (0,100)
        start time
        test with Triangulation
        stop time
        value=start-stop
        record value in list
    average of values in list
    record avr value in new list
results of test
```

Table II show the results recorded during test. Test was done on the same environment (same system). We did this test for polygons with 20 and 180 sides. It was observed that increment of sides would just lead to increment of time. Results are showing how long each test ran, times $10^{-5}$ seconds.

Test results were showing almost same results with little variation. Fig.10 and Fig.11 clearly demonstrates this, so outcomes were reliable, hence open for quantitative discussion.

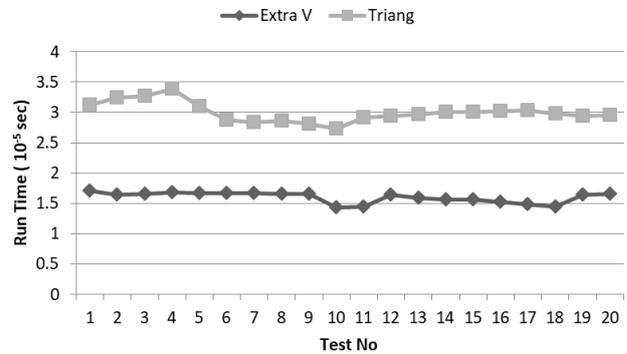

Fig. 10. Results for polygon with 20 sides.

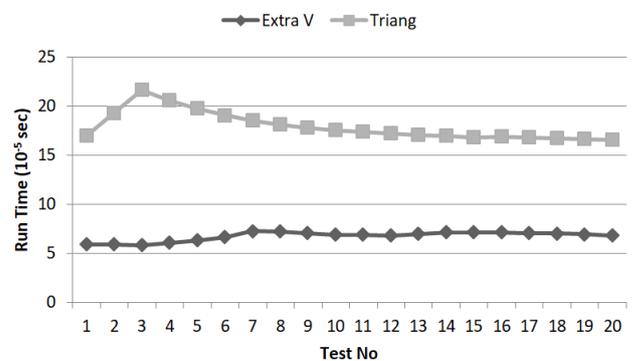

Fig. 11. Results for polygon with 180 sides.

## IV. RESULTS AND DISCUSSIONS

The results shown in Table 2 clearly demonstrate that extra vertex concept is much more effective than ordinary triangulation method. With increment of number of vertices of the polygon, extra vertex becoming much more powerful tool comparing to old method.

Table II

| Test | 20 vertices | | 180 vertices | |
| --- | --- | --- | --- | --- |
| | ExtaV. | Trian. | ExtraV. | Trian. |
| 1 | 1.71 | 3.12 | 5.89 | 17.02 |
| 2 | 1.64 | 3.24 | 5.95 | 19.32 |
| 3 | 1.66 | 3.27 | 5.81 | 21.64 |
| 4 | 1.68 | 3.389 | 6.05 | 20.58 |
| 5 | 1.67 | 3.1 | 6.36 | 19.78 |
| 6 | 1.67 | 2.88 | 6.69 | 19.07 |
| 7 | 1.67 | 2.83 | 7.27 | 18.57 |
| 8 | 1.66 | 2.86 | 7.26 | 18.16 |
| 9 | 1.65 | 2.81 | 7.06 | 17.84 |
| 10 | 1.43 | 2.73 | 6.93 | 17.59 |
| 11 | 1.45 | 2.92 | 6.91 | 17.37 |
| 12 | 1.64 | 2.94 | 6.83 | 17.25 |
| 13 | 1.59 | 2.97 | 7.02 | 17.1 |
| 14 | 1.56 | 3.01 | 7.16 | 16.97 |
| 15 | 1.56 | 3.01 | 7.16 | 16.85 |
| 16 | 1.52 | 3.02 | 7.16 | 16.89 |
| 17 | 1.49 | 3.03 | 7.09 | 16.8 |
| 18 | 1.45 | 2.98 | 7.03 | 16.72 |
| 19 | 1.64 | 2.94 | 6.94 | 16.68 |
| 20 | 1.66 | 2.96 | 6.81 | 16.61 |

This paper has discussed application of the new approach to Shoelace algorithm usage in area calculation in convex polygons. As it was mentioned above, for the case when the polygon in concave, this method needs some additions; still same algorithm of calculation can be used to calculate area, but some extra conditions should be also evaluated. But this left for further researches. Yet, there are already some existing algorithms to do calculation for concave polygons, like Ray Shooting method.

## V. ACKNOWLEFGEMENT

I would like to thank my supervisor Prof. Nwojo Agwu, Nile University of Nigeria, for his support and guidance during the experiment and paper preparation.